\documentclass[a4paper,11pt]{article}
\usepackage{amsmath}
\usepackage{amsfonts}
\usepackage{amssymb}
\usepackage{bm}
\usepackage[utf8x]{inputenc}
\usepackage{ae}
\usepackage[T1]{fontenc}
\usepackage[english]{babel}
\usepackage{breakurl}
\usepackage{caption}
\usepackage{csquotes}
\usepackage{xcolor}
\usepackage{graphicx}

\usepackage{makecell}
\usepackage{placeins}
\usepackage{tabularx}
\usepackage{tabulary}
\usepackage{setspace}
\usepackage{tablefootnote}
\usepackage[flushmargin]{footmisc}
\usepackage[comma]{natbib}
\usepackage{array}
\usepackage{float}
\usepackage{caption}
\usepackage{tikz}
\usetikzlibrary{positioning,arrows.meta}
\usetikzlibrary{shapes.geometric}
\usepackage[a4paper,left=2.5cm,right=2.5cm,top=3cm,bottom=3cm]{geometry}
\usepackage[colorinlistoftodos]{todonotes}
\usepackage{pdfpages}
\usepackage{longtable}
\usepackage{adjustbox}
\usepackage{booktabs}
\usepackage{multirow}
\usepackage{lscape}
\usepackage[colorlinks=true, allcolors=blue]{hyperref}
\hypersetup{breaklinks=true}

\usepackage[flushleft]{threeparttable}
\usepackage{inputenc}

\usepackage[left]{lineno}



\begin{document} \doublespacing \pagestyle{plain}
	
	
	\def\ci{\perp\!\!\!\perp}
	\begin{center}
		
		{\LARGE Does fare communication format matter? Evidence from a randomised experiment in public transport}

		{\large \vspace{0.8cm}}
		
		{\large Hannes Wallimann, Silvio Sticher, and Christian Weibel}\medskip

		{\small {University of Applied Sciences and Arts Lucerne, Institute of Tourism and Mobility} \smallskip }
		
		{\large \vspace{0.8cm}}
		
		{\large 2026}\smallskip

	\end{center}
	
	\smallskip

\noindent \textbf{Abstract:} {Highly concentrated demand in public transport places pressure on infrastructure and reduces service quality. Pricing instruments---such as discounted off-peak single fare tickets---are a widely used tool for rescheduling demand across time. Yet no study has directly examined whether the price communication format of single fare tickets affects travel behaviour in terms of departure-time rescheduling. To fill this gap, we exploit a randomised stated preference survey conducted in Switzerland based on participants' actual journeys ($n = 1{,}082$), assigning participants to one of three price formats: absolute amounts, absolute deviations, or percentage deviations from their current fare. Estimating a multinomial logit model with price--format interaction terms, we find no significant effect of communication format on aggregate price sensitivity. However, for low-fare trips, percentage-based communication substantially amplifies price sensitivity relative to absolute amounts or absolute deviations. Moreover, when prices are shown as percentage deviations, price sensitivity increases for small price spreads and decreases for large price spreads, with a crossover point corresponding to a discount depth of approximately 47\%. In summary, these findings suggest that percentage-based price communication has the potential to reinforce the demand-shifting effect of off-peak discounts, particularly for low-fare products and for discounts shallower than approximately 47\%.}
	
	{\small \smallskip }
	{\small \smallskip }
	{\small \smallskip }
	
	
	{\small \smallskip }
	{\small \smallskip }
	{\small \smallskip }
	
	{\small \noindent \textbf{Acknowledgments:} We are indebted to Luca Binswanger, Moritz Blattner, Lukas Egloff, Johannes Josi, and Seït Käch for helpful discussions and support. Moreover, we are grateful to Alliance Swisspass for financial support.}
	
	\bigskip
	\bigskip
	\bigskip
	\bigskip
	
	{\small {\scriptsize 
\begin{spacing}{1.5}\noindent  
\textbf{Addresses for correspondence:} Hannes Wallimann,  \href{mailto:hannes.wallimann@hslu.ch}{hannes.wallimann@hslu.ch}; Silvio Sticher, \href{mailto:silvio.sticher@hslu.ch}{silvio.sticher@hslu.ch}; Christian Weibel, \href{mailto:christian.weibel@hslu.ch}{christian.weibel@hslu.ch}.
\end{spacing}
			
		}\thispagestyle{empty}\pagebreak  }

	{\small \renewcommand{\thefootnote}{\arabic{footnote}} %
		\setcounter{footnote}{0}  \pagebreak \setcounter{footnote}{0} \pagebreak %
		\setcounter{page}{1} }
	
\section{Introduction}\label{introduction}

Highly concentrated demand is a major concern of public transport 
authorities and operators, as it places pressure on infrastructure 
and reduces service quality \citep[see, e.g.,][]{li2011crowding, 
	lu2024analyzing}: the discomfort of passengers 
is higher in a crowded vehicle than in one with available seats 
\citep{batarce2016valuing}. Policies that aim at redistributing passengers by inducing departure-time rescheduling are therefore of central interest in the transport literature. Such policies range from infrastructure investments and operational measures such as increased service frequency to demand-side pricing instruments. The latter include studies on peak pricing \citep{halvorsen2016reducing,ancupane2026peak} 
and off-peak pricing \citep{huber2022business,thommen2023price} 
of single fare tickets.

Single fare ticket prices can, however, be presented to 
passengers in more than one way, and shifts in the reference 
point against which options are evaluated can alter preferences, 
as cognitive judgments are influenced by the manner in which 
decision problems are framed \citep[e.g.,][]{kahneman1979prospect, 
	kahneman1984choices}. In particular, providing an explicit 
reference point can make an otherwise hard-to-evaluate attribute 
easier to assess and thus increase the weight it receives in the 
choice \citep{hsee1996evaluability,thaler1985mental}. In practice, public transport companies must decide how to communicate price differences: as absolute amounts, as monetary deviations 
from the regular fare, or as percentage deviations. To the best 
of our knowledge, no study has directly examined whether the price 
communication format of single fare tickets affects public 
transport behaviour in terms of travel scheduling. 

To fill this gap, we exploit a randomised stated-preference survey 
conducted alongside a public transport field test in Switzerland, 
with 1{,}082 participants (out of 2{,}910 contacted; response rate: 39.6\%). The survey 
consists of eight choice tasks, each based on a trip randomly 
drawn from the respondent's own travel history in the weeks prior 
to the survey. In each task, respondents chose between two 
hypothetical alternatives or opted out of travelling by public 
transport altogether. The central manipulation concerns the 
presentation of the price attribute, to which respondents were 
randomly assigned: (i)~\textit{Absolute Amount}, where the fare is 
shown as an absolute CHF amount and respondents must construct 
their own reference point; (ii)~\textit{Absolute Deviation}, 
where the fare is shown as an absolute monetary deviation from 
the respondent's actual fare; and (iii)~\textit{Relative 
	Deviation}, where the fare is shown as a percentage deviation 
from the respondent's actual fare.

We estimate a multinomial logit model with price--version 
interaction terms, exploiting the randomised assignment to 
identify the causal effect of price communication format on 
mode choice behaviour. At the aggregate level, respondents 
exposed to deviation-based price formats do not exhibit 
significantly higher price sensitivity than those exposed to 
the Absolute Amount format. However, our results indicate that price communication format affects travel behaviour through two sources of heterogeneous treatment effects. First, for low fares, the Relative Deviation format substantially amplifies price sensitivity relative to the Absolute Amount format or Absolute Deviation format, rendering passengers more responsive to price differences and therefore more likely to reschedule their trip by selecting the cheaper alternative---consistent with the Rule of 100 \citep{chen1998effects,gonzalez2016amount}. Second, when prices are shown as percentage deviations, price sensitivity is stronger for small price spreads and weaker for large ones, with a crossover at a discount depth of approximately 47\%.

The remainder of the paper proceeds as follows. 
Section~\ref{sec:theory} develops the theoretical background 
and derives the hypotheses. Section~\ref{sec:methods} describes 
the survey instrument, the experimental design, and the 
econometric model. Section~\ref{sec:results} presents the 
results. Finally, Section~\ref{sec:discussion} concludes and 
discusses the results.

\section{Theoretical Background and Hypotheses}\label{sec:theory}

Our survey presents single-ticket choices in three different versions: as absolute amounts (e.g. "CHF\,59.20"), as absolute deviations (e.g., "CHF\,15.95 more expensive"), or as relative deviations (e.g., "37\% more expensive"). This section develops theoretical arguments for how price presentation format may affect price sensitivity in single fare ticket choices, and derives testable hypotheses.

\subsection{Price Format and Price Sensitivity}\label{sec:theory:priceFormat}

Cognitive judgments are systematically influenced by how a decision problem is framed \citep{kahneman1979prospect,kahneman1984choices,tversky1991loss}. Two mechanisms are particularly important in this regard: i) reference dependence, i.e., the standard against which an option is evaluated, and ii) loss aversion, i.e., the steeper value function for losses than for gains \citep{tversky1991loss}. Price increases therefore affect consumers subjectively more strongly than equivalent price decreases \citep{mcgraw2010comparing,novemsky2005boundaries}.

Evidence from the transport literature supports reference dependence for fares. Using a large-scale stated preference dataset of Sydney car drivers, \citet{hess2008asymmetric} find evidence that respondents react asymmetrically to attribute increases versus decreases relative to a reference alternative, and that ignoring this asymmetry leads to substantially biased preference parameter estimates. \citet{stathopoulos2012revisiting} show in a stated choice study of UK rail and bus commuters that fare evaluations depend on the reference point. Against the actual fare, responses are largely symmetric, whereas against respondents acceptable or ideal fares, increases (i.e. losses) weigh more than decreases (i.e. gains).

The reference point must also be constructed in stated preference tasks. Several studies show that reference point framing systematically alters how consumers evaluate pay-per-use costs \citep[e.g., ][]{yun2022consumer}, a finding with direct relevance to usage-based public transport tariffs \citep{weibel2024choosing}. \citet{rose2008designing} show that when presenting attribute levels in stated preference experiments, participants must internally construct their own reference point against which to evaluate the alternative. Moreover, when attribute levels are expressed as deviations from a respondent's reference trip (the so-called pivot design), price changes become more salient and, consequently, increase the perceived disutility of price increases.

The deviation versions state the change relative to the respondent's current fare. In line with the evaluability hypothesis \citep{hsee1996evaluability}, providing a comparison value can make an otherwise hard to judge attribute easier to evaluate and thus receive greater weight in the choice decision. This is also consistent with transaction utility theory \citep{thaler1985mental}, according to which the value of a purchase depends not only on what is received for the price paid but also on how that price compares with a reference price.

In our setting, the three questionnaire versions differ in how they set the reference point. The Absolute Amount version presents no explicit deviation---respondents must construct their own reference point. By contrast, the Absolute Deviation and Relative Deviation versions explicitly anchor evaluation to the current myRIDE fare (see Section~\ref{sec:survey}), making price changes more salient and easier to evaluate. In summary, the mechanisms discussed above suggest that deviation-based formats increase the weight respondents place on price. This yields our first hypothesis:

\label{hyp:h1}
\begin{quote}
	\textbf{H1:} \textit{Respondents exposed to deviation-based price formats (Absolute or Relative Deviation) exhibit higher price sensitivity than those exposed to the Absolute Amount format.}
\end{quote}

\subsection{Fare Level Effects}\label{Theory:Level}

Fare levels vary substantially across trips, from short urban journeys to longer interregional connections. Whether deviation-based formats alter price sensitivity may therefore depend on the fare level itself. A well-documented example is the so-called “Rule of 100” \citep[see, e.g., ][]{berger2013contagious}: below a price threshold of approximately 100 USD, relative differences loom larger than absolute ones; above it, the relationship reverses \citep{gonzalez2016amount}. Applied to our setting, this suggests that the Relative Deviation format---which communicates price changes as percentages---should be particularly salient for low-fare connections, where the same percentage maps onto a small absolute amount, and comparatively less so for high-fare connections.

Several mechanisms are discussed for why the format changes how large a difference is perceived. First, the numerosity heuristic suggests that the format yielding the larger number appears larger \citep{pelham1994easy}. Second, relative thinking implies that a fixed amount matters less against a high base \citep{darke1993deciding,tversky1981framing}, and consumers also tend to neglect the base value when processing percentage information \citep{chen2012more}. Third, on a perceptual level, the Weber--Fechner law states that whether the difference between two stimuli is noticed 
is not determined by the absolute difference, but by the 
ratio to the initial reference \citep{monroe1971psychophysics}.

In summary, we expect that these mechanisms jointly predict that the percentage format is more influential where it yields the larger number, i.e., at low price levels. In our setting, fares in the experiment are usage-based and vary substantially across trips---from short urban to long-distance connections. This fare variation within the same experiment gives rise to heterogeneous treatment effects: the effect of the price communication format on price sensitivity is not constant across observations, but depends on the actual fare level faced by the respondent. Due to the Rule of 100 and relative thinking, the Relative Deviation format in particular should amplify perceived price changes more strongly at low fare levels, where a given percentage deviation translates into a small absolute amount that would otherwise appear negligible. This yields our next hypothesis:\footnote{Note that not all empirical evidence points in the same direction. \citet{gendall2006message} find little or no framing effect for low-priced goods but an advantage of absolute framing for high-priced goods. \citet{hardesty2003consumer} report a preference for percentage presentation when the promotional benefit level is high, but not when the promotional benefit level is moderate. \citet{lehtimaki2019influence} find that the attractiveness of absolute discounts varies with the regular price level, whereas percentage discounts are evaluated more consistently across price levels. We follow \citet{chen1998effects} and the Rule of 100 \citep{berger2013contagious,gonzalez2016amount} in predicting that the percentage format dominates for low-price goods, while acknowledging that the empirical evidence is mixed.}

\begin{quote}
	\textbf{H2:}\label{hyp:h2} \textit{The framing effect of the Relative Deviation format on price sensitivity is stronger for low fares than for high fares.}
\end{quote}

\subsection{Price Spread and the Relative Deviation Format}\label{sec:theory:priceSpread}

When designing dynamic single-fare pricing (whether through off-peak 
discounts or peak surcharges) operators must set not only the fare 
level but also the size of the price spread between options. Beyond 
its absolute size, the format in which this spread is communicated 
may modulate how salient it appears to the decision-maker.

Applied to price spread, the same logic as in 
Section~\ref{Theory:Level} applies: a given percentage differential 
between two fare options communicates less variation in relative terms 
the larger that spread already is. Under the Relative Deviation 
format, which presents the price difference as a percentage, the 
communicated spread therefore shrinks in relative terms as the actual 
spread grows. Under the Absolute Amount or Absolute Deviation formats, 
by contrast, a larger spread remains directly visible as a larger 
number and retains its salience.

This yields our second hypothesis:

\begin{quote}
	\textbf{H3:}\label{hyp:h3} \textit{The increase in price sensitivity as price spread grows is weaker under the Relative Deviation format than under non-relative formats.}
\end{quote}

\section{Methods and experimental setting}\label{sec:methods}

\subsection{Survey Instrument}\label{sec:survey}

The survey was distributed by the Swiss Federal Railways (SBB) to all customers participating in the myRIDE field test, a distance-based, progressive public transport tariff system field test launched by Alliance SwissPass in Switzerland in 2024 \citep{wallimann2025effect}. Under myRIDE, fares depend on the total number of tariff kilometres travelled per day, with volume-dependent discounts applied on a monthly basis. Trips (and the personalised fare history) were recorded automatically via a smartphone application. 

Respondents were informed that the survey was conducted by the Lucerne University of Applied Sciences and Arts (HSLU) on behalf of Alliance SwissPass and SBB. Participation was voluntary, and respondents were informed that their answers would be used anonymously for scientific purposes. Completion of the questionnaire took approximately 5 to 10 minutes according to the communication to the participants (on average it took about 6 minutes). As an incentive, participants were entered into a draw for a public transport voucher worth CHF~1,000. The questionnaire was administered in German and French, with respondents selecting their preferred language at the outset. A total of 2{,}910 individuals were contacted, of whom 1{,}152 completed the survey, yielding a response rate of 39.6\%. After excluding respondents who did not fully complete the questionnaire or reported implausible age values, the final estimation sample amounts to 1{,}082 respondents.

The core of the survey consists of eight stated-preference (SP) choice tasks, each based on a trip drawn randomly with replacement from the respondent's own revealed-preference data in the weeks prior to the survey. Within each task, respondents were shown two hypothetical alternatives to either the outbound or return leg of their journey, alongside their actual trip. The direction shown (outbound or return) was randomised across tasks. Respondents were asked to choose between Option~A and Option~B (both hypothetical alternatives), or to indicate that they would not have made the trip by public transport if only these two options would have been available (see also sensitivity analysis in Section \ref{section:sensitivity}).

Each choice task presented four attributes for both hypothetical alternatives (as well as for the actual trip): departure time, travel duration (in minutes), mode chain, and price. The mode chain refers to the sequence of public transport modes used (e.g., Regional Train~-- Intercity Train). The hypothetical alternatives were generated as follows. Departure time was shifted by a fixed offset drawn randomly from the set $\{-120, -60, -45, -30, -20, -10, +10, +20, +30, +45, +60, +120\}$ minutes, with the constraint that the two alternatives received distinct offsets. Travel duration and price were each scaled by a factor drawn independently and uniformly from $[0.5, 2.0]$ relative to the actual trip values (see also Figure \ref{fig:price_distribution} in Appendix \ref{Appendix_Figures} for the price distribution). The mode chain was drawn randomly from the set of chains observed across both the actual connection and the available alternative connections for the respondent in a narrow time window around the departure of the actual trip. To ensure choice task validity, any combination in which one alternative strictly dominated the other (i.e., shorter travel time, lower price, and identical mode chain simultaneously) was discarded and redrawn.

The central manipulation of this study concerns how the price attribute was presented to respondents. Participants were randomly assigned to one of three questionnaire versions, which differed exclusively in the presentation of the price attribute:

\begin{itemize}
	\item \textbf{Absolute Amount (AA, 50\% of the participants):} Price 	is shown as an absolute CHF amount (e.g., "CHF~12.50"). Respondents 	must internally construct their own reference point based on the current fare.
	
	\item \textbf{Absolute Deviation (AbsDev, 25\% of the participants):} Price is shown as an absolute monetary deviation from the respondent's actual myRIDE fare for that trip, expressed as a verbal label (e.g., ``CHF~3.50 cheaper'' or ``CHF~2.02 more expensive''). The current fare serves as an explicit reference 
	point.
	
	\item \textbf{Relative Deviation (RelDev, 25\% of the participants):} Price is shown as a percentage deviation from the respondent's actual myRIDE fare, again expressed as a verbal label (e.g., ``35\% more expensive'' or ``20\% cheaper''). The current fare serves as an explicit reference point, but the deviation is 
	expressed in proportional terms.
\end{itemize}

Table~\ref{tab:sp_example} illustrates a fictitious choice task for each of the three versions. All other attributes (departure time, travel duration, and mode chain) were communicated identical across versions. For the purpose of model estimation, prices in the AbsDev and RelDev versions were converted back to absolute CHF amounts using the respondent's actual myRIDE fare as the reference (Section~\ref{sec:model}).

\begin{table}[ht]
	\centering
	\caption{Two illustrative choice alternatives across the three price formats}
	\label{tab:sp_example}
	\small
	\begin{tabular}{lcccc}
		\toprule
		& & \multicolumn{3}{c}{Price attribute by version} \\
		\cmidrule(lr){3-5}
		Attribute & Actual Trip & AA & AbsDev & RelDev \\
		\midrule
		Departure time         & 08:00     & 06:00     & 06:00                    & 06:00               \\
		Travel duration (min.) & 60        & 39        & 39                       & 39                  \\
		Mode chain             & IC        & IC        & IC                       & IC                  \\
		Price                  & CHF~43.25 & CHF~59.20 & CHF~15.95 more expensive & 37\% more expensive \\
		\midrule
		Departure time         & 08:00     & 08:20     & 08:20                    & 08:20               \\
		Travel duration (min.) & 60        & 84        & 84                       & 84                  \\
		Mode chain             & IC        & IR~--~IC & IR~--~IC & IR~--~IC\\
		Price                  & CHF~43.25 & CHF~27.00 & CHF~16.25 cheaper        & 38\% cheaper        \\
		\bottomrule
		\multicolumn{5}{l}{\footnotesize \textit{Notes:} Two illustrative choice alternatives from a fictitious choice task.} \\
		\multicolumn{5}{l}{\footnotesize IC = InterCity; IR = InterRegio (shown in full in the experiment).} \\
	\end{tabular}
\end{table}

Following the SP tasks, respondents provided sociodemographic information including age, gender, household size, highest educational attainment, number of cars in the household, possession of a driving licence, and monthly gross household income (in CHF).

\subsection{Identification Strategy}\label{sec:identification}

We adopt the potential outcomes framework of 
\citet{rubin1974estimating} to formalise the identification of 
framing effects on stated preference estimates. Let 
$D \in \{\text{AA}, \text{AbsDev}, \text{RelDev}\}$ denote the 
questionnaire version randomly assigned to respondent $i$, where AA 
(Absolute Amount) serves as the reference condition, and let 
$Y_i(D)$ denote the potential choice outcome of respondent $i$ under 
version $D$. We consider three parameters of interest. The first is 
the version-specific price coefficient $\beta_{\text{price}}(D)$, 
which addresses Hypothesis~\hyperref[hyp:h1]{H1}. The second is the 
interaction between the price coefficient and a binary low-fare 
indicator ($\ell_i = \mathbf{1}[\text{fare}_i \leq \text{median 
	fare}]$), equal to one if the respondent's actual myRIDE reference 
fare falls below the sample median of CHF~11.93, and zero otherwise; 
this addresses Hypothesis~\hyperref[hyp:h2]{H2}. The third is the 
interaction between the price coefficient and the relative price 
spread between options, defined as $\Delta p = |p_A - p_B| / 
\bar{p}_{AB}$ where $\bar{p}_{AB} = (p_A + p_B)/2$, consistent with 
Hypothesis~\hyperref[hyp:h3]{H3}. For estimation, $\Delta p$ is 
centred at its sample mean: $\Delta p_c = \Delta p - 
\overline{\Delta p}$.

Identification relies on two assumptions. First, the \textbf{Independence Assumption} requires that questionnaire version assignment is statistically independent of respondents' background characteristics and unobserved heterogeneity:

\begin{equation}
	\{Y(D)\}_{D \in \mathcal{D}} \perp D
	\label{eq:independence}
\end{equation}

\noindent This assumption is satisfied by design: questionnaire versions were randomly assigned to participants during the field test, with no self-selection into treatment. Looking at the balance table shown in Table~\ref{tab:balance}, we see that with the exception of a marginally significant difference in the share of above-median income households between the Absolute Deviation and Relative Deviation groups ($p = 0.029$), and a statistically significant difference in mean fare levels across versions that reflects variation in respondents' actual myRIDE reference fares rather than experimental design asymmetry, no systematic differences in observed characteristics are detected across the three versions.

Second, the \textbf{Stable Unit Treatment Value Assumption} (SUTVA) requires that the stated choices of respondent $i$ are unaffected by the questionnaire version assigned to other respondents. Given that respondents completed the 
survey independently and were unaware of the existence of alternative questionnaire versions, SUTVA seems plausible.

\subsection{Econometric Model}\label{sec:model}

In the baseline specification, we model a standard multinomial logit (MNL) model that specifies the utility of alternative $j$ for individual $i$ as:

\begin{equation}
	U_{ij} = \beta_{\text{price}} \cdot p_{ij} + \beta_{\text{time}} \cdot t_{ij} + \boldsymbol{\gamma}' \mathbf{x}_{ij} + \varepsilon_{ij}
	\label{eq:utility_base}
\end{equation}

\noindent where $p_{ij}$ is the fare, $t_{ij}$ is the travel time, and $\mathbf{x}_{ij}$ is a vector of trip attributes including the number of transfers and mode chain dummies (long-distance, local, other; reference: regional). We extend this baseline specification in three steps to test the hypotheses developed in Section~\ref{sec:theory}.

First, we introduce two dummy variables identifying the questionnaire version assigned to each respondent:

\begin{equation}
	D_{\text{abs}} = \begin{cases} 1 & \text{if Absolute Deviation} \\ 0 & \text{otherwise} \end{cases}
	\qquad
	D_{\text{rel}} = \begin{cases} 1 & \text{if Relative Deviation} \\ 0 & \text{otherwise} \end{cases}
	\label{eq:dummies}
\end{equation}

\noindent Second, we extend this baseline in three steps, introducing 
version dummies $D_{\text{abs}}$ and $D_{\text{rel}}$ (see 
Equation~\ref{eq:dummies}), low-fare interactions ($\ell_i$, 
Hypothesis~\hyperref[hyp:h2]{H2}, and price-spread interactions ($\Delta p_c$, 
Hypothesis~\hyperref[hyp:h3]{H3}, as defined in Section~\ref{sec:identification}. 
The full utility function is:\footnote{The Absolute Deviation 
	interaction term $\beta_{\text{abs}}$ is excluded following a 
	cluster-robust Wald test ($p = 0.73$); see 
	Section~\ref{sec:results}. Nested specifications are compared using cluster-robust Wald tests.}

\begin{equation}
	\begin{aligned}
		U_{ij} = \Bigl(&\beta_{\text{price}}
		+ \beta_{\text{rel}} \cdot D_{\text{rel}}
		+ \beta_{\ell} \cdot \ell_i
		+ \beta_{\ell \times \text{rel}} \cdot \ell_i \cdot D_{\text{rel}} \\
		&+ \beta_{\Delta p} \cdot \Delta p_c
		+ \beta_{\Delta p \times \text{rel}} \cdot \Delta p_c \cdot D_{\text{rel}}
		\Bigr) \cdot p_{ij}
		+ \beta_{\text{time}} \cdot t_{ij}
		+ \boldsymbol{\gamma}' \mathbf{x}_{ij}
		+ \varepsilon_{ij}
	\end{aligned}
	\label{eq:utility}
\end{equation}

\noindent The model is estimated by maximum likelihood using the \texttt{mlogit} package in R \citep{croissant2020mlogit}, with standard errors clustered at the respondent level to account for dependence between choice tasks.

\section{Results}\label{sec:results}

\subsection{Price Format and Price Sensitivity}\label{Section:priceFormat}

\begin{table}[H]
	\centering
	\caption{Baseline Model: Framing Effects on Price Sensitivity}
	\label{tab:baseline}
	\small
	\renewcommand{\arraystretch}{0.85}
	\begin{tabular}{lcccc}
		\toprule
		& \multicolumn{2}{c}{(1) Baseline} & \multicolumn{2}{c}{(2) + Trip Attributes} \\
		\cmidrule(lr){2-3} \cmidrule(lr){4-5}
		Variable & Est. & SE & Est. & SE \\
		\midrule
		\multicolumn{5}{l}{\textit{Price coefficients}} \\
		$\beta_{\text{price}}$                       & $-$0.0938$^{***}$ & 0.0072 & $-$0.0953$^{***}$ & 0.0072 \\
		$\beta_{\text{abs}}$ (Price $\times$ $D_{\text{abs}}$) & $-$0.0051 & 0.0121 & $-$0.0042         & 0.0122 \\
		$\beta_{\text{rel}}$ (Price $\times$ $D_{\text{rel}}$) & $+$0.0104 & 0.0113 & $+$0.0106         & 0.0114 \\
		\multicolumn{5}{l}{\textit{Time coefficient}} \\
		$\beta_{\text{time}}$ (Travel time, min)     & $-$0.0350$^{***}$ & 0.0018 & $-$0.0353$^{***}$ & 0.0018 \\
		\multicolumn{5}{l}{\textit{Trip attributes ($\boldsymbol{\gamma}$)}} \\
		Transfers                                    & ---               & ---    & $-$0.3389$^{***}$ & 0.0550 \\
		Mode: long-distance                          & ---               & ---    & $+$0.1076         & 0.0659 \\
		Mode: local                                  & ---               & ---    & $-$0.2799$^{*}$   & 0.1152 \\
		Mode: other                                  & ---               & ---    & $-$0.7887         & 0.5276 \\
		\midrule
		\multicolumn{5}{l}{\textit{Model fit}} \\
		Log-likelihood          & \multicolumn{2}{l}{$-$3{,}813.96} & \multicolumn{2}{l}{$-$3{,}784.78} \\
		Wald-test vs.\ previous & \multicolumn{2}{l}{---}            & \multicolumn{2}{l}{$\chi^2(4) = 55.51$, $p < 0.001$} \\
		Observations            & \multicolumn{2}{l}{6{,}528}        & \multicolumn{2}{l}{6{,}528} \\
		\bottomrule
		\multicolumn{5}{l}{\footnotesize \textit{Notes:} Standard errors clustered at the respondent level ($n = 1{,}082$).} \\
		\multicolumn{5}{l}{\footnotesize Absolute Amount (AA) serves as the reference category. Mode reference: regional.} \\
		\multicolumn{5}{l}{\footnotesize $^{*}p < 0.05$, $^{**}p < 0.01$, $^{***}p < 0.001$.} \\
	\end{tabular}
\end{table}

Table~\ref{tab:baseline} reports the results of the baseline
model. The price coefficient $\hat{\beta}_{\text{price}}$ and
the travel time coefficient $\hat{\beta}_{\text{time}}$ point
in the expected direction across both specifications, confirming
that respondents trade off cost and time in the experiment.
Adding the number of transfers and mode chain dummies in
Model~(2) improves model fit significantly
($\chi^2(4) = 55.51$, $p < 0.001$). However, the framing
interaction terms $\beta_{\text{abs}}$ and $\beta_{\text{rel}}$
remain statistically indistinguishable from zero in both
specifications, failing to support
Hypothesis~\hyperref[hyp:h1]{H1}. Put differently,
deviation-based formats do not alter the weight respondents
place on price relative to the Absolute Amount format.

\subsection{Fare Level Effects}\label{Section:Fare}

Before turning to the fare level models, we verify that
the Absolute Deviation interaction term $\beta_{\text{abs}}$
can be excluded from all subsequent specifications. A
cluster-robust Wald test fails to reject the null hypothesis
$H_0: \beta_{\text{abs}} = 0$ ($p = 0.73$), and the remaining
coefficients are virtually unaffected by its omission. All
models from this point therefore (and to be consistent with Section \ref{Theory:Level}) retain only the Relative
Deviation interaction $\beta_{\text{rel}}$.

As discussed in Section \ref{Theory:Level}, consumers perceive relative price differences as larger below a price threshold, while absolute differences loom
larger above it. Applied to our setting, this suggests that
the Relative Deviation format---which communicates discounts
as percentages---should be particularly salient for
low-fare connections, where the same percentage deviation
maps onto a small absolute amount. For high-fare connections,
by contrast, absolute amounts are likely to dominate
perception regardless of the presentation format, attenuating
any framing effect.

\begin{table}[H]
	\centering
	\caption{Fare Level Effects}
	\label{tab:level}
	\small
	\renewcommand{\arraystretch}{0.85}
	\begin{tabular}{lcccccc}
		\toprule
		& \multicolumn{2}{c}{(1) Baseline} & \multicolumn{2}{c}{(2) + $\ell_i$} & \multicolumn{2}{c}{(3) + $\ell_i \times D_{\text{rel}}$ (Final)} \\
		\cmidrule(lr){2-3} \cmidrule(lr){4-5} \cmidrule(lr){6-7}
		Variable & Est. & SE & Est. & SE & Est. & SE \\
		\midrule
		\multicolumn{7}{l}{\textit{Price coefficients}} \\
		$\beta_{\text{price}}$                                                       & $-$0.0967$^{***}$ & 0.0064 & $-$0.0795$^{***}$ & 0.0055 & $-$0.0809$^{***}$ & 0.0057 \\
		$\beta_{\text{rel}}$ (Price $\times$ $D_{\text{rel}}$)                       & $+$0.0120         & 0.0108 & $+$0.0051         & 0.0095 & $+$0.0106         & 0.0096 \\
		$\beta_{\ell}$ (Price $\times$ $\ell_i$)                                     & ---               & ---    & $-$0.1498$^{***}$ & 0.0129 & $-$0.1329$^{***}$ & 0.0138 \\
		$\beta_{\ell \times \text{rel}}$ (Price $\times$ $\ell_i$ $\times$ $D_{\text{rel}}$) & ---       & ---    & ---               & ---    & $-$0.1004$^{**}$  & 0.0374 \\
		\multicolumn{7}{l}{\textit{Time coefficient}} \\
		$\beta_{\text{time}}$ (Travel time, min)                                     & $-$0.0353$^{***}$ & 0.0018 & $-$0.0356$^{***}$ & 0.0018 & $-$0.0355$^{***}$ & 0.0018 \\
		\multicolumn{7}{l}{\textit{Trip attributes ($\boldsymbol{\gamma}$)}} \\
		Transfers                                                                    & $-$0.3390$^{***}$ & 0.0550 & $-$0.3326$^{***}$ & 0.0557 & $-$0.3310$^{***}$ & 0.0556 \\
		Mode: long-distance                                                          & $+$0.1071         & 0.0659 & $+$0.1037         & 0.0682 & $+$0.1028         & 0.0681 \\
		Mode: local                                                                  & $-$0.2811$^{*}$   & 0.1153 & $-$0.3153$^{*}$   & 0.1230 & $-$0.3181$^{**}$  & 0.1229 \\
		Mode: other                                                                  & $-$0.7902         & 0.5265 & $-$0.8670         & 0.5351 & $-$0.8557         & 0.5271 \\
		\midrule
		\multicolumn{7}{l}{\textit{Model fit}} \\
		Log-likelihood          & \multicolumn{2}{l}{$-$3{,}784.94} & \multicolumn{2}{l}{$-$3{,}667.13} & \multicolumn{2}{l}{$-$3{,}660.91} \\
		Wald-test vs.\ previous & \multicolumn{2}{l}{---}            & \multicolumn{2}{l}{$\chi^2(1) = 134.35$, $p < 0.001$} & \multicolumn{2}{l}{$\chi^2(1) = 7.21$, $p < 0.01$} \\
		Observations            & \multicolumn{2}{l}{6{,}528}        & \multicolumn{2}{l}{6{,}528}        & \multicolumn{2}{l}{6{,}528} \\
		\midrule
		\multicolumn{7}{l}{\textit{Effective price coefficients at $\overline{\Delta p}_c = 0$}} \\
		AA, high fare ($\ell_i = 0$)      & \multicolumn{2}{l}{$-$0.0967} & \multicolumn{2}{l}{$-$0.0795} & \multicolumn{2}{l}{$-$0.0809} \\
		AA, low fare ($\ell_i = 1$)       & \multicolumn{2}{l}{---}        & \multicolumn{2}{l}{$-$0.2293} & \multicolumn{2}{l}{$-$0.2138} \\
		RelDev, high fare ($\ell_i = 0$)  & \multicolumn{2}{l}{$-$0.0847} & \multicolumn{2}{l}{$-$0.0744} & \multicolumn{2}{l}{$-$0.0703} \\
		RelDev, low fare ($\ell_i = 1$)   & \multicolumn{2}{l}{---}        & \multicolumn{2}{l}{$-$0.2242} & \multicolumn{2}{l}{$-$0.3035} \\
		\bottomrule
		\multicolumn{7}{l}{\footnotesize \textit{Notes:} Standard errors clustered at the respondent level ($n = 1{,}082$).} \\
		\multicolumn{7}{l}{\footnotesize AA and AbsDev serve as the joint reference category. Mode reference: regional.} \\
		\multicolumn{7}{l}{\footnotesize $\ell_i = \mathbf{1}[\mathrm{fare}_i \leq \mathrm{median\;fare}]$: binary low-fare indicator (median: CHF~11.93).} \\
		\multicolumn{7}{l}{\footnotesize $^{*}p < 0.05$, $^{**}p < 0.01$, $^{***}p < 0.001$.} \\
	\end{tabular}
\end{table}

Table~\ref{tab:level} reports the results. Model~(1)
corresponds to the trip-attribute specification of
Model~(2) from Table~\ref{tab:baseline}, now restricted
to exclude the Absolute Deviation interaction term.
Adding the pooled low-fare interaction $\beta_{\ell}$
in Model~(2) substantially improves model fit
($\chi^2(1) = 134.35$, $p < 0.001$), confirming that
fare level is a driver of price sensitivity.
The format-specific interaction $\beta_{\ell \times \text{rel}}$
in Model~(3) provides a further significant improvement
($\chi^2(1) = 7.21$, $p < 0.01$), and constitutes the
specification on which the subsequent price spread
analysis builds.

For all communication formats, respondents are
substantially more price-sensitive for low-fare trips
than for high-fare trips ($\hat{\beta}_{\ell} = -0.1329$,
$p < 0.001$). At the mean price differential
($\overline{\Delta p}_c = 0$), the effective price
coefficient for low fares under the Absolute Amount
format is $-$0.214, compared to $-$0.081 for high
fares---a factor of approximately 2.6. For instance,
a CHF~5 price increase on a low-fare connection
(e.g., from CHF~5 to CHF~10) is substantially more
likely to induce respondents to switch to a cheaper
alternative than an equivalent absolute increase on a
high-fare connection (e.g., from CHF~50 to CHF~55).

Moreover, the interaction term
$\hat{\beta}_{\ell \times \text{rel}} = -0.1004$
($p = 0.007$) is negative and statistically significant,
consistent with Hypothesis~\hyperref[hyp:h2]{H2} that percentage frames are most salient for low-price goods. Under
the Relative Deviation format, the effective price
coefficient for low fares reaches $-$0.304---more than
four times larger in absolute value than the corresponding
high-fare coefficient ($-$0.070)---while for high fares
the format difference is negligible ($-$0.070 vs.\
$-$0.081).

\subsection{Price Spread and the Relative Deviation Format}\label{Section:PriceDifferences}

\begin{table}[H]
	\centering
	\caption{Price Differential Magnitude and Framing Effects}
	\label{tab:magnitude}
	\footnotesize
	\renewcommand{\arraystretch}{0.85}
	\begin{tabular}{lcccccc}
		\toprule
		& \multicolumn{2}{c}{(1) Baseline} & \multicolumn{2}{c}{(2) $+\Delta p_c$} & \multicolumn{2}{c}{(3) $+\Delta p_c \times D_{\text{rel}}$ (Final)} \\
		\cmidrule(lr){2-3} \cmidrule(lr){4-5} \cmidrule(lr){6-7}
		Variable & Est. & SE & Est. & SE & Est. & SE \\
		\midrule
		\multicolumn{7}{l}{\textit{Price coefficients}} \\
		$\beta_{\text{price}}$                                       & $-$0.0809$^{***}$ & 0.0057 & $-$0.0959$^{***}$ & 0.0079 & $-$0.0891$^{***}$ & 0.0082 \\
		$\beta_{\text{rel}}$ \quad(Price $\times$ $D_{\text{rel}}$) & $+$0.0106         & 0.0096 & $+$0.0100         & 0.0090 & $-$0.0175         & 0.0125 \\
		$\beta_{\ell}$ \quad(Price $\times$ $\ell_i$)               & $-$0.1329$^{***}$ & 0.0138 & $-$0.1340$^{***}$ & 0.0134 & $-$0.1339$^{***}$ & 0.0136 \\
		$\beta_{\ell \times \text{rel}}$ \quad(Price $\times$ $\ell_i \times D_{\text{rel}}$) & $-$0.1004$^{**}$ & 0.0374 & $-$0.0975$^{**}$ & 0.0365 & $-$0.0932$^{**}$ & 0.0358 \\
		$\beta_{\Delta p}$ \quad(Price $\times$ $\Delta p_c$)       & ---               & ---    & $+$0.0529$^{***}$ & 0.0131 & $+$0.0267$^{.}$   & 0.0155 \\
		$\beta_{\Delta p \times \text{rel}}$ \quad(Price $\times$ $\Delta p_c \times D_{\text{rel}}$) & --- & --- & --- & --- & $+$0.0915$^{***}$ & 0.0254 \\
		\multicolumn{7}{l}{\textit{Time coefficient}} \\
		$\beta_{\text{time}}$ \quad(Travel time, min)                & $-$0.0355$^{***}$ & 0.0018 & $-$0.0366$^{***}$ & 0.0019 & $-$0.0368$^{***}$ & 0.0019 \\
		\multicolumn{7}{l}{\textit{Trip attributes ($\boldsymbol{\gamma}$)}} \\
		Transfers          & $-$0.3310$^{***}$ & 0.0556 & $-$0.3385$^{***}$ & 0.0561 & $-$0.3417$^{***}$ & 0.0561 \\
		Mode: long-dist.   & $+$0.1028         & 0.0681 & $+$0.1070         & 0.0681 & $+$0.1081         & 0.0683 \\
		Mode: local        & $-$0.3181$^{**}$  & 0.1229 & $-$0.3241$^{**}$  & 0.1229 & $-$0.3182$^{**}$  & 0.1227 \\
		Mode: other        & $-$0.8557         & 0.5271 & $-$0.8958         & 0.5571 & $-$0.8571         & 0.5457 \\
		\midrule
		\multicolumn{7}{l}{\textit{Model fit}} \\
		Log-likelihood          & \multicolumn{2}{l}{$-$3{,}660.91} & \multicolumn{2}{l}{$-$3{,}649.44} & \multicolumn{2}{l}{$-$3{,}641.94} \\
		Wald-test vs.\ previous & \multicolumn{2}{l}{---}            & \multicolumn{2}{l}{$\chi^2(1) = 16.36$, $p < 0.001$} & \multicolumn{2}{l}{$\chi^2(1) = 13.03$, $p < 0.001$} \\
		Observations            & \multicolumn{2}{l}{6{,}528}        & \multicolumn{2}{l}{6{,}528}        & \multicolumn{2}{l}{6{,}528} \\
		\midrule
		\multicolumn{7}{l}{\textit{Loss aversion tests (vs.\ Model~3)}} \\
		Wald: $+$ price increase dummy ($D_{\text{up}}$)                       & \multicolumn{6}{l}{$p = 0.125$} \\
		Wald: $+$ price increase $\times$ $D_{\text{rel}}$ ($D_{\text{up}} \times D_{\text{rel}}$) & \multicolumn{6}{l}{$p = 0.333$} \\
		\bottomrule
		\multicolumn{7}{l}{\footnotesize \textit{Notes:} Standard errors clustered at the respondent level ($n = 1{,}082$).} \\
		\multicolumn{7}{l}{\footnotesize AA and AbsDev serve as the joint reference category. Mode reference: regional.} \\
		\multicolumn{7}{l}{\footnotesize $\Delta p_c = \Delta p - \overline{\Delta p}$, where $\Delta p = |p_A - p_B|/\bar{p}_{AB}$, centred at sample mean $\overline{\Delta p} = 0.42$.} \\
		\multicolumn{7}{l}{\footnotesize $\ell_i = \mathbf{1}[\mathrm{fare}_i \leq \mathrm{median\;fare}]$: binary low-fare indicator (median: CHF~11.93).} \\
		\multicolumn{7}{l}{\footnotesize $^{.}p < 0.10$, $^{*}p < 0.05$, $^{**}p < 0.01$, $^{***}p < 0.001$.} \\
	\end{tabular}
\end{table}

Table~\ref{tab:magnitude} reports the results of the price
spread models, building on the fare level specification
(Model~3 of Table~\ref{tab:level}), which serves as the
baseline here.

Adding the main effect of price spread $\beta_{\Delta p}$
in Model~(2) significantly improves model fit
($\chi^2(1) = 16.36$, $p < 0.001$). The coefficient itself
is positive but only marginally significant
($\hat{\beta}_{\Delta p} = +0.0254$, $p = 0.086$),
suggesting that a wider price gap between options reduces
price sensitivity on average across formats---but this
pooled effect is primarily driven by the RelDev group,
as the interaction in Model~(3) reveals.

\begin{figure}[H]
	\centering
	\caption{Relative Framing and Price Sensitivity: Effect of Price Spread}
	\label{fig:diff_plot}
	\scalebox{0.9}{
\begin{tikzpicture}[x=1pt,y=1pt]
\definecolor{fillColor}{RGB}{255,255,255}
\path[use as bounding box,fill=fillColor,fill opacity=0.00] (0,0) rectangle (505.89,289.08);
\begin{scope}
\path[clip] (  0.00,  0.00) rectangle (505.89,289.08);
\definecolor{fillColor}{RGB}{255,255,255}

\path[fill=fillColor] (  0.00,  0.00) rectangle (505.89,289.08);
\end{scope}
\begin{scope}
\path[clip] ( 41.49, 30.69) rectangle (500.39,283.58);
\definecolor{drawColor}{gray}{0.92}

\path[draw=drawColor,line width= 0.6pt,line join=round] ( 41.49, 33.00) --
	(500.39, 33.00);

\path[draw=drawColor,line width= 0.6pt,line join=round] ( 41.49, 83.23) --
	(500.39, 83.23);

\path[draw=drawColor,line width= 0.6pt,line join=round] ( 41.49,133.46) --
	(500.39,133.46);

\path[draw=drawColor,line width= 0.6pt,line join=round] ( 41.49,183.70) --
	(500.39,183.70);

\path[draw=drawColor,line width= 0.6pt,line join=round] ( 41.49,233.93) --
	(500.39,233.93);

\path[draw=drawColor,line width= 0.6pt,line join=round] ( 62.35, 30.69) --
	( 62.35,283.58);

\path[draw=drawColor,line width= 0.6pt,line join=round] (166.64, 30.69) --
	(166.64,283.58);

\path[draw=drawColor,line width= 0.6pt,line join=round] (270.94, 30.69) --
	(270.94,283.58);

\path[draw=drawColor,line width= 0.6pt,line join=round] (375.23, 30.69) --
	(375.23,283.58);

\path[draw=drawColor,line width= 0.6pt,line join=round] (479.53, 30.69) --
	(479.53,283.58);
\definecolor{drawColor}{RGB}{238,106,135}

\path[draw=drawColor,line width= 1.1pt,line join=round] ( 62.35, 42.18) --
	( 66.52, 44.48) --
	( 70.69, 46.78) --
	( 74.86, 49.08) --
	( 79.03, 51.38) --
	( 83.21, 53.68) --
	( 87.38, 55.98) --
	( 91.55, 58.27) --
	( 95.72, 60.57) --
	( 99.89, 62.87) --
	(104.07, 65.17) --
	(108.24, 67.47) --
	(112.41, 69.77) --
	(116.58, 72.07) --
	(120.75, 74.37) --
	(124.92, 76.67) --
	(129.10, 78.97) --
	(133.27, 81.26) --
	(137.44, 83.56) --
	(141.61, 85.86) --
	(145.78, 88.16) --
	(149.96, 90.46) --
	(154.13, 92.76) --
	(158.30, 95.06) --
	(162.47, 97.36) --
	(166.64, 99.66) --
	(170.81,101.96) --
	(174.99,104.26) --
	(179.16,106.55) --
	(183.33,108.85) --
	(187.50,111.15) --
	(191.67,113.45) --
	(195.85,115.75) --
	(200.02,118.05) --
	(204.19,120.35) --
	(208.36,122.65) --
	(212.53,124.95) --
	(216.70,127.25) --
	(220.88,129.54) --
	(225.05,131.84) --
	(229.22,134.14) --
	(233.39,136.44) --
	(237.56,138.74) --
	(241.74,141.04) --
	(245.91,143.34) --
	(250.08,145.64) --
	(254.25,147.94) --
	(258.42,150.24) --
	(262.60,152.53) --
	(266.77,154.83) --
	(270.94,157.13) --
	(275.11,159.43) --
	(279.28,161.73) --
	(283.45,164.03) --
	(287.63,166.33) --
	(291.80,168.63) --
	(295.97,170.93) --
	(300.14,173.23) --
	(304.31,175.53) --
	(308.49,177.82) --
	(312.66,180.12) --
	(316.83,182.42) --
	(321.00,184.72) --
	(325.17,187.02) --
	(329.34,189.32) --
	(333.52,191.62) --
	(337.69,193.92) --
	(341.86,196.22) --
	(346.03,198.52) --
	(350.20,200.81) --
	(354.38,203.11) --
	(358.55,205.41) --
	(362.72,207.71) --
	(366.89,210.01) --
	(371.06,212.31) --
	(375.23,214.61) --
	(379.41,216.91) --
	(383.58,219.21) --
	(387.75,221.51) --
	(391.92,223.81) --
	(396.09,226.10) --
	(400.27,228.40) --
	(404.44,230.70) --
	(408.61,233.00) --
	(412.78,235.30) --
	(416.95,237.60) --
	(421.13,239.90) --
	(425.30,242.20) --
	(429.47,244.50) --
	(433.64,246.80) --
	(437.81,249.09) --
	(441.98,251.39) --
	(446.16,253.69) --
	(450.33,255.99) --
	(454.50,258.29) --
	(458.67,260.59) --
	(462.84,262.89) --
	(467.02,265.19) --
	(471.19,267.49) --
	(475.36,269.79) --
	(479.53,272.08);
\definecolor{drawColor}{gray}{0.40}

\path[draw=drawColor,line width= 0.6pt,dash pattern=on 4pt off 4pt ,line join=round] ( 41.49,183.70) -- (500.39,183.70);

\path[draw=drawColor,line width= 0.6pt,dash pattern=on 1pt off 3pt ,line join=round] (319.14, 30.69) -- (319.14,283.58);
\definecolor{drawColor}{gray}{0.60}

\path[draw=drawColor,line width= 0.5pt,line join=round] (239.26, 30.69) -- (239.26,283.58);
\definecolor{drawColor}{RGB}{0,0,0}

\node[text=drawColor,anchor=base west,inner sep=0pt, outer sep=0pt, scale=  0.85] at (327.49,260.31) {Crossover: $\Delta p = $0.62};
\definecolor{drawColor}{gray}{0.40}

\node[text=drawColor,anchor=base west,inner sep=0pt, outer sep=0pt, scale=  0.85] at (247.61, 53.39) {$\overline{\Delta p}$};
\end{scope}
\begin{scope}
\path[clip] (  0.00,  0.00) rectangle (505.89,289.08);
\definecolor{drawColor}{gray}{0.30}

\node[text=drawColor,anchor=base east,inner sep=0pt, outer sep=0pt, scale=  0.88] at ( 36.54, 29.97) {-0.06};

\node[text=drawColor,anchor=base east,inner sep=0pt, outer sep=0pt, scale=  0.88] at ( 36.54, 80.20) {-0.04};

\node[text=drawColor,anchor=base east,inner sep=0pt, outer sep=0pt, scale=  0.88] at ( 36.54,130.43) {-0.02};

\node[text=drawColor,anchor=base east,inner sep=0pt, outer sep=0pt, scale=  0.88] at ( 36.54,180.67) {0.00};

\node[text=drawColor,anchor=base east,inner sep=0pt, outer sep=0pt, scale=  0.88] at ( 36.54,230.90) {0.02};
\end{scope}
\begin{scope}
\path[clip] (  0.00,  0.00) rectangle (505.89,289.08);
\definecolor{drawColor}{gray}{0.30}

\node[text=drawColor,anchor=base,inner sep=0pt, outer sep=0pt, scale=  0.88] at ( 62.35, 19.68) {0.00};

\node[text=drawColor,anchor=base,inner sep=0pt, outer sep=0pt, scale=  0.88] at (166.64, 19.68) {0.25};

\node[text=drawColor,anchor=base,inner sep=0pt, outer sep=0pt, scale=  0.88] at (270.94, 19.68) {0.50};

\node[text=drawColor,anchor=base,inner sep=0pt, outer sep=0pt, scale=  0.88] at (375.23, 19.68) {0.75};

\node[text=drawColor,anchor=base,inner sep=0pt, outer sep=0pt, scale=  0.88] at (479.53, 19.68) {1.00};
\end{scope}
\begin{scope}
\path[clip] (  0.00,  0.00) rectangle (505.89,289.08);
\definecolor{drawColor}{RGB}{0,0,0}

\node[text=drawColor,anchor=base,inner sep=0pt, outer sep=0pt, scale=  1.10] at (270.94,  7.64) {Relative price spread $\Delta p = |p_A - p_B|/\bar{p}_{AB}$};
\end{scope}
\begin{scope}
\path[clip] (  0.00,  0.00) rectangle (505.89,289.08);
\definecolor{drawColor}{RGB}{0,0,0}

\node[text=drawColor,rotate= 90.00,anchor=base,inner sep=0pt, outer sep=0pt, scale=  1.10] at ( 13.08,157.13) {Marginal effect of RelDev on price sensitivity};
\end{scope}
\end{tikzpicture}}
	\begin{minipage}{0.9\textwidth}
		\footnotesize \textit{Notes:} The y-axis shows the marginal
		effect of the Relative Deviation format on price sensitivity,
		as a function of the relative price spread
		$\Delta p = |p_A - p_B|/\bar{p}_{AB}$. Negative values
		indicate that RelDev respondents are more price-sensitive
		than AA respondents; positive values indicate lower price
		sensitivity. The dotted vertical line indicates the crossover
		at $\Delta p \approx 0.62$; the solid vertical line indicates
		the sample mean $\overline{\Delta p} = 0.42$.
	\end{minipage}
\end{figure}
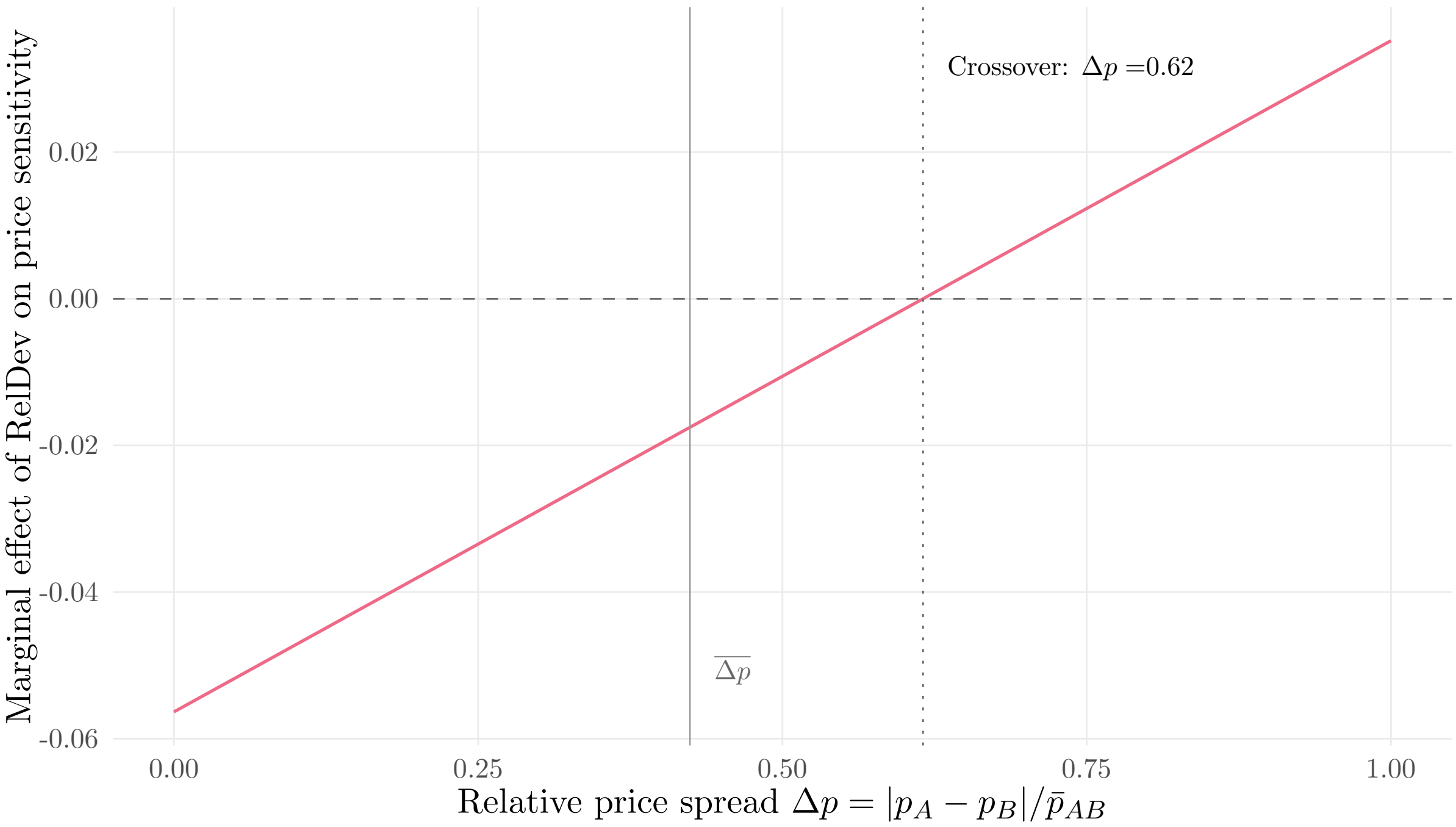

The interaction between price spread and the Relative
Deviation format $\beta_{\Delta p \times \text{rel}}$ in
Model~(3) is positive and significant
($+$0.0915, $p < 0.001$), supporting
Hypothesis~\hyperref[hyp:h3]{H3}. When prices are
presented as percentage deviations, larger price
differentials between options reduce price sensitivity
relative to the Absolute Amount format: the Relative
Deviation format subjects the communicated spread to
diminishing perceptual sensitivity, dampening its impact
on choice. Figure~\ref{fig:diff_plot} illustrates this
pattern: below a crossover point of $\Delta p \approx 0.62$,
RelDev respondents are more price-sensitive
than respondents shown the absolute fare or absolute deviation,
while above it the relationship reverses., while above it the relationship reverses.
This crossover is directly relevant for the design of
discounted single fare tickets: for large discounts
exceeding approximately 47\% (i.e., $\Delta p > 0.62$),
presenting the price reduction in relative terms reduces
passengers' price sensitivity relative to showing absolute
amounts.

Finally, augmenting Model~(3) with a dummy for price
increases relative to the actual myRIDE fare---and its
interaction with $D_{\text{rel}}$---leaves model fit
virtually unchanged ($p = 0.125$ and $p = 0.333$,
respectively), suggesting that the spread effect is
driven by the presentation format and the magnitude
of the price differential rather than by asymmetric
responses to price increases versus decreases
(see loss aversion tests at the bottom of
Table~\ref{tab:magnitude}).\footnote{\citet{hess2008asymmetric},
	using a large-scale stated preference dataset of Sydney car
	drivers, find clear evidence that respondents react
	asymmetrically to attribute increases versus decreases
	relative to a reference alternative. In our setting, such
	asymmetry does not appear to play a role at the 5\%
	significance level.}

\subsection{Sensitivity Analysis}\label{section:sensitivity}

In the following, we discuss two potential concerns regarding 
our analysis.

First, opt-out behaviour across questionnaire versions may 
confound the connection choice estimates. If RelDev respondents 
with large price differentials opt out at higher rates than AA 
respondents, the remaining sample would be systematically less 
price-sensitive in the RelDev group, biasing the framing effect 
estimates. To assess this, we estimate a logistic regression of 
the opt-out indicator on the relative price spread $\Delta p$, 
questionnaire version, and their interaction, restricting the 
sample to the Absolute Amount and Relative Deviation versions. A 
likelihood ratio test of the null hypothesis of parallel opt-out 
gradients fails to reject ($\chi^2 = 0.568$, df~$= 1$, 
$p = 0.451$), indicating that the relationship between price 
spread and opt-out behaviour does not differ between the two 
versions. Figure~\ref{fig:optout} in 
Appendix~\ref{Appendix_Figures} illustrates the parallel 
predicted opt-out gradients.

Second, approximately 11.3\% of respondents ($n = 122$) 
consistently selected the same option across all eight choice 
tasks, suggesting potential attribute non-attendance. 
Table~\ref{tab:robustness} in Appendix~\ref{Appendix_A} 
compares the key coefficients of the final model estimated on 
the full sample and on the trimmed sample excluding these 
respondents. All coefficients remain stable in sign and 
magnitude: the largest relative change is observed for 
$\hat{\beta}_{\text{rel}}$ ($-$0.0175 vs.\ $-$0.0126), which 
remains statistically insignificant in both specifications. The 
fare level and price spread interaction terms 
$\hat{\beta}_{\ell}$, $\hat{\beta}_{\ell \times \text{rel}}$, 
and $\hat{\beta}_{\Delta p \times \text{rel}}$ are virtually 
unchanged, confirming that the main results are not driven by 
respondents with potential attribute non-attendance.

\section{Discussion and Conclusion}\label{sec:discussion}

\subsection{Discussion of findings}

In our paper, we examined whether the price communication format of single 
fare tickets affects public transport behaviour in terms of travel 
scheduling. To this end, we analysed a randomised stated preference 
survey with 1{,}082 participants in Switzerland. In eight choice 
tasks, each respondent was shown two hypothetical alternatives 
constructed from their personal travel history. The experimental 
manipulation concerned the presentation of the price attribute: 
respondents were randomly assigned to one of three conditions: the 
absolute fare level in Swiss francs, the absolute deviation from their 
current fare, or the percentage deviation from their current fare.

At the aggregate level, price communication format did not 
significantly affect respondents' price sensitivity. 
This null result could be interpreted as a rejection of our arguments in Section \ref{sec:theory:priceFormat}, underscoring an opposing view that for frequently-used products, such as public transport tickets, lead consumers to form an internal 
reference price against which future prices are evaluated 
\citep{mazumdar2005reference,lin2017follow}. Drawing on adaptation-level theory, this 
internal reference price can be understood as a judgment standard 
shaped by prior price experience \citep{helson1964adaptation,monroe1973buyers,thomas2007internal}, 
creating stable expectations that serve as reference points for new 
price information \citep{lattin1989reference}. If such an internal 
reference point is already well-established, providing an explicit 
external reference may add little incremental information, thereby 
attenuating the treatment effect on aggregate price sensitivity. 

However, the results reveal two sources of heterogeneous treatment effects. First, as discussed in Section~\ref{Section:Fare}, for low-fare trips, percentage-based price communication substantially amplifies price sensitivity, rendering respondents more responsive to price differences and therefore more likely to select the cheaper alternative. This finding is consistent with the Rule of 100 and related empirical work showing that relative price differences are perceived as larger when the base price is small \citep{chen1998effects,gonzalez2016amount}, and supports Hypothesis~\hyperref[hyp:h2]{H2}.

Second, as discussed in Section~\ref{Section:PriceDifferences}, when
prices are presented as percentage deviations, the size of the price
spread between options modulates price sensitivity: smaller spreads
amplify price sensitivity relative to the Absolute Amount format,
while larger spreads reduce it. Below a crossover point of
$\Delta p \approx 0.62$ (where $\Delta p$ denotes the price
differential between the two options divided by their mean),
respondents shown the percentage difference are more price-sensitive
than respondents shown the absolute fare or absolute deviation,
while above it the relationship reverses. This crossover pattern is consistent with
Hypothesis~\hyperref[hyp:h3]{H3}.

The reversal above the crossover---where larger spreads reduce price
sensitivity under the Relative Deviation format relative to the
Absolute Amount format---warrants further explanation. Percentage
labels make the depth of a price difference immediately visible,
whereas the Absolute Amount format requires it to be inferred.
Research on discounts suggests that consumers perceive gains as
smaller than advertised \citep{gupta1992discounting}, and that the
effect of discount depth levels off beyond a saturation point
\citep{gupta1992discounting,mazumdar2005reference}, potentially
triggering scepticism at large magnitudes \citep{darke2005effects}.
This is related to models with attribute cutoffs, in which marginal
(dis)utility changes beyond a certain threshold \citep{swait2001non}.
Note, however, that these findings are based on stated purchase
intentions in response to single advertised discounts
\citep[e.g.,][]{gupta1992discounting}, not on spreads between options
in a choice setting, and are therefore not directly comparable to our
results.

\subsection{Implications for fare communication}\label{sec:implications}

To contextualise the magnitude of these effects, we turn to the Swiss public transport market, where the experiment was conducted. In 2025, Swiss public transport generated total revenues of 
CHF 7.04 billion.\footnote{See \url{https://www.allianceswisspass.ch/de/asp/News/Newsmeldung?filterCategory=4-20&newsid=997}, accessed on 15 October 2026.} Single fare tickets, however, account for only 
31\% of this revenue. A further 20\% is attributable to the 
so-called 'General Abonnement' (GA), a season ticket that 
provides unlimited travel for one month or one year across virtually 
the entire Swiss public transport network. In addition, Switzerland's 
20 regional tariff associations offer annual and monthly season 
tickets with comparable features to the GA, but restricted to their 
respective regions or designated fare zones. The remaining revenue derives from these regional passes and day tickets, which likewise offer no (or limited) scope for 
pricing-based demand management. Pricing-based demand management effects can therefore mainly be 
achieved through the 31\% of journeys sold as single fare tickets. In 
Switzerland, so-called supersaver tickets offer discounts of up to 70\% on selected, primarily long-distance routes. The volume of such discounted tickets is regulated through 
a mutual agreement with the Swiss price watchdog: for instance, the Swiss Federal 
Railways are required to sell supersaver tickets worth at 
least CHF 90 million in long-distance services between 
1 January and 31 December 2027.\footnote{See 
	\url{https://www.admin.ch/de/newnsb/AU8_APWrkLN5RPPMOlCGk}, 
	accessed on 15 October 2026.} Given that the crossover identified 
in our analysis occurs at $\Delta p = 0.62$---corresponding to a 
discount depth of approximately 47\%---only the highest discounts on single fare tickets exceed this threshold. For the majority of 
discounted single fare tickets, percentage-based price 
communication therefore retains its price-sensitivity-amplifying 
effect. This effect is expected to be particularly pronounced for 
low-fare products.

Against the background of heavily concentrated peak-hour demand---and its associated pressure 
on infrastructure and service quality---operators may therefore 
have an incentive to reconsider the current emphasis on unlimited-use 
subscriptions in favour of more flexible, trip-based pricing schemes 
that allow for targeted demand management. Note that the structural constraint of a (relatively) small proportion of single fare tickets is, however, specific to the Swiss 
context, even if comparable unlimited-use subscription schemes exist 
elsewhere. Examples include the Klimaticket in Austria \citep[see, e.g., ][]{wallimann2024austria} 
or the Deutschlandticket in Germany \citep[see, e.g., ][]{waldorf2025low}, where single 
and flexible-fare tickets may account for a substantially larger 
share of revenue. In the United Kingdom, for instance, season tickets accounted for only 14\% of public service operator ticket sales in the second quarter of 2026, while single and flexible-fare tickets made up the remaining 86\% \citep{orr2026passenger}. In markets with this 
ticket-type structure, the potential for percentage-based price 
communication is correspondingly larger.

Moreover, note that the transport literature suggests that peak and off-peak pricing can effectively shift travel demand. For Switzerland, 
\citet{huber2022business} estimate that a one percentage point 
increase in discounts raises rescheduling propensity by 0.16 
percentage points even among customers travelling without a 
discount---implying that rescheduling, while achievable, comes at 
a non-trivial cost. The same study finds that leisure travellers are 
substantially more responsive to discounts than commuters. Also other studies find effect heterogeneity, e.g., simulation studies suggest that off-peak discounts are more effective in reducing afternoon than morning peaks \citep{lovric2016evaluating,adnan2020examining}, as commuters tend to shift the evening rather than the morning trip \citep{adnan2020examining}. An 
open question for future research is whether such heterogeneity 
also moderates the framing effects documented in the present study---for instance, whether leisure travellers, being more price-responsive, are also more susceptible to percentage-based 
price communication.

Finally, note that frequent price reductions may raise 
doubts about product quality, potentially undermining the 
effectiveness of discount strategies \citep{darke2005effects}. 

\subsection{Limitations and further research}

Two limitations should be noted. First, the sample is 
conditional on voluntary participation in the myRIDE field test. 
Participants were required to install a smartphone application and 
to actively engage with a novel, technology-based fare system, 
suggesting that they are likely more tech-savvy and more frequent 
public transport users than the general population. For such 
experienced users, internal reference prices may be particularly 
well-established, which could attenuate the framing effects 
documented here relative to what might be observed among less 
experienced travellers. More broadly, the generalisability of the 
findings to other public transport systems (or even other industries)---which may differ 
substantially in ticket-type composition, fare elasticities, and 
passengers' willingness to pay---remains an open question for 
future research.

Second, due to the stated preference experimental setting, 
hypothetical bias cannot be excluded: respondents may behave 
differently in real booking situations than in hypothetical choice 
tasks. A related concern is the plausibility of 
experimentally generated fares. Prices perceived as implausible are 
unlikely to shift valuations, as consumers tend to reject price 
information that falls outside the range anchored by their internal 
reference price \citep{lin2017follow,mazumdar2005reference}. According to assimilation-contrast theory, new price information is only integrated into the internal reference price, if it is judged to belong to the range of acceptable prices; prices outside of this range are not assimilated and therefore carry little weight in the evaluation \citep{sherif1961social,mazumdar2005reference}. Consumers with short repurchase cycles rely more strongly on memory-based than on new external price information \citep{mazumdar2005reference}, and repeated price evaluations increase confidence in one's own price knowledge \citep{thomas2007internal}. This 
rejection is expected to be stronger for frequently purchased 
products, where repeated experience stabilises the internal 
reference price and reduces the acceptance range for external price 
signals \citep{thomas2007internal}. In the present study, 
experimentally generated fares were constructed by scaling the 
respondent's actual myRIDE fare by a factor drawn from $[0.5, 
2.0]$, implying that extreme deviations represent a minority of 
observations (see Figure~\ref{fig:price_distribution}). 
Nevertheless, the possibility that some respondents discounted 
implausible price levels cannot be fully ruled out.


\newpage

\section*{Declaration of generative AI and AI-assisted technologies in the manuscript preparation process}

We used large language models (mainly Claude by Anthropic and Consensus) as supportive tools for programming assistance, literature screening and linguistic editing of the manuscript. All outputs were carefully reviewed and verified by the authors, who take responsibility for the content presented.

\newpage
\bigskip

\bibliographystyle{sageh}
\bibliography{myRide_price.bib}

\newpage
\bigskip

\begin{appendix}
	
	\numberwithin{equation}{section}
	\counterwithin{figure}{section}
	\counterwithin{table}{section}
	\noindent \textbf{\LARGE Appendices}
	
	\section{Additional Tables}\label{Appendix_A}

	\begin{table}[htbp]
		\centering
		\caption{Balance Table by Questionnaire Version}
		\label{tab:balance}
		\small
		\begin{tabularx}{\textwidth}{lXXXXXX}
			\toprule
			& \multicolumn{3}{c}{Mean (SD)} & \multicolumn{3}{c}{$p$-value (Welch $t$-test)} \\
			\cmidrule(lr){2-4} \cmidrule(lr){5-7}
			Variable & AA & AbsDev & RelDev & AA vs.\ AbsDev & AA vs.\ RelDev & AbsDev vs.\ RelDev \\
			\midrule
			$N$ & 538 & 288 & 272 & & & \\
			\multicolumn{7}{l}{\textit{Sociodemographics}} \\
			Age (years)               & \shortstack[l]{42.6\\(13.0)} & \shortstack[l]{43.3\\(13.1)} & \shortstack[l]{42.3\\(12.5)} & 0.436 & 0.793 & 0.359 \\[8pt]
			Female (share)            & \shortstack[l]{0.3\\(0.5)}   & \shortstack[l]{0.3\\(0.5)}   & \shortstack[l]{0.3\\(0.5)}   & 0.414 & 0.967 & 0.457 \\[8pt]
			Driving licence (share)   & \shortstack[l]{0.9\\(0.3)}   & \shortstack[l]{0.9\\(0.3)}   & \shortstack[l]{0.9\\(0.3)}   & 0.698 & 0.633 & 0.933 \\[8pt]
			University degree (share) & \shortstack[l]{0.6\\(0.5)}   & \shortstack[l]{0.6\\(0.5)}   & \shortstack[l]{0.6\\(0.5)}   & 0.102 & 0.848 & 0.212 \\[8pt]
			Household size            & \shortstack[l]{2.4\\(1.1)}   & \shortstack[l]{2.4\\(1.2)}   & \shortstack[l]{2.4\\(1.2)}   & 0.851 & 0.876 & 0.768 \\[8pt]
			Car ownership (share)     & \shortstack[l]{0.6\\(0.5)}   & \shortstack[l]{0.6\\(0.5)}   & \shortstack[l]{0.7\\(0.5)}   & 0.948 & 0.272 & 0.365 \\[8pt]
			Above median income       & \shortstack[l]{0.6\\(0.5)}   & \shortstack[l]{0.7\\(0.5)}   & \shortstack[l]{0.6\\(0.5)}   & 0.493 & 0.074 & 0.029$^{*}$ \\[8pt]
			\multicolumn{7}{l}{\textit{Experiment attributes}} \\
			Fare (CHF)                & \shortstack[l]{20.7\\(20.6)} & \shortstack[l]{19.7\\(20.0)} & \shortstack[l]{21.1\\(20.5)} & 0.017$^{*}$ & 0.417 & 0.006$^{**}$ \\[8pt]
			Travel time (min)         & \shortstack[l]{47.1\\(40.5)} & \shortstack[l]{46.6\\(41.8)} & \shortstack[l]{47.5\\(40.5)} & 0.616 & 0.581 & 0.367 \\
			\bottomrule
			\multicolumn{7}{l}{\footnotesize \textit{Notes:} AA = Absolute Amount; AbsDev = Absolute Deviation; RelDev = Relative Deviation.} \\
			\multicolumn{7}{l}{\footnotesize Mean and standard deviation (in parentheses). Above median income: $\geq$ CHF~8,000/month.} \\
			\multicolumn{7}{l}{\footnotesize $p$-values from two-sided Welch $t$-tests. $^{*}p < 0.05$, $^{**}p < 0.01$. Shares coded as binary (0/1).} \\
			\multicolumn{7}{l}{\footnotesize The significant fare difference reflects variation in respondents' actual myRIDE reference} \\
			\multicolumn{7}{l}{\footnotesize fares across versions, not experimental design asymmetry.} \\
			\multicolumn{7}{l}{\footnotesize Respondents with implausible age values excluded ($n = 3$).} \\
		\end{tabularx}
	\end{table}
	
	\begin{table}[H]
		\centering
		\caption{Robustness Check: Excluding Always-Same Respondents ($n = 122$)}
		\label{tab:robustness}
		\small
		\renewcommand{\arraystretch}{0.85}
		\begin{tabular}{lcccc}
			\toprule
			& \multicolumn{2}{c}{Full sample ($n = 1{,}082$)} & \multicolumn{2}{c}{Trimmed ($n = 960$)} \\
			\cmidrule(lr){2-3} \cmidrule(lr){4-5}
			Parameter & Est. & SE & Est. & SE \\
			\midrule
			\multicolumn{5}{l}{\textit{Price coefficients}} \\
			$\beta_{\text{price}}$                                          & $-$0.0891$^{***}$ & 0.0082 & $-$0.0878$^{***}$ & 0.0085 \\
			$\beta_{\text{rel}}$ (Price $\times$ RelDev)                    & $-$0.0175         & 0.0125 & $-$0.0126         & 0.0127 \\
			$\beta_{\Delta p}$ (Price $\times$ $\Delta p_c$)                & $+$0.0267$^{.}$   & 0.0155 & $+$0.0230         & 0.0159 \\
			$\beta_{\Delta p \times \text{rel}}$ (Price $\times$ $\Delta p_c$ $\times$ RelDev) & $+$0.0915$^{***}$ & 0.0254 & $+$0.0909$^{***}$ & 0.0259 \\
			$\beta_{\ell}$ (Price $\times$ $\ell$)                          & $-$0.1339$^{***}$ & 0.0136 & $-$0.1320$^{***}$ & 0.0139 \\
			$\beta_{\ell \times \text{rel}}$ (Price $\times$ $\ell$ $\times$ RelDev) & $-$0.0932$^{**}$ & 0.0358 & $-$0.0972$^{**}$  & 0.0360 \\
			\multicolumn{5}{l}{\textit{Time coefficient}} \\
			$\beta_{\text{time}}$ (Travel time, min)                        & $-$0.0368$^{***}$ & 0.0019 & $-$0.0360$^{***}$ & 0.0019 \\
			\multicolumn{5}{l}{\textit{Trip attributes ($\boldsymbol{\gamma}$)}} \\
			Transfers                                                       & $-$0.3417$^{***}$ & 0.0561 & $-$0.3404$^{***}$ & 0.0577 \\
			Mode: long-distance                                             & $+$0.1081         & 0.0683 & $+$0.1124         & 0.0690 \\
			Mode: local                                                     & $-$0.3182$^{**}$  & 0.1227 & $-$0.3456$^{**}$  & 0.1258 \\
			Mode: other                                                     & $-$0.8571         & 0.5457 & $-$0.7969         & 0.4999 \\
			\midrule
			\multicolumn{5}{l}{\textit{Model fit}} \\
			Log-likelihood & \multicolumn{2}{l}{$-$3{,}641.94} & \multicolumn{2}{l}{$-$3{,}460.89} \\
			\bottomrule
			\multicolumn{5}{l}{\footnotesize \textit{Notes:} Standard errors clustered at the respondent level.} \\
			\multicolumn{5}{l}{\footnotesize Trimmed sample excludes 122 respondents who consistently selected} \\
			\multicolumn{5}{l}{\footnotesize the same option across all eight choice tasks (attribute non-attendance).} \\
			\multicolumn{5}{l}{\footnotesize AA and AbsDev serve as the joint reference category. Mode reference: regional.} \\
			\multicolumn{5}{l}{\footnotesize $^{.}p < 0.10$, $^{*}p < 0.05$, $^{**}p < 0.01$, $^{***}p < 0.001$.} \\
		\end{tabular}
	\end{table}
	
	\section{Additional Figures}\label{Appendix_Figures}

	\begin{figure}[H]
		\centering
		\caption{Distribution of Experimental Fares by Questionnaire Version}
		\label{fig:price_distribution}
		\resizebox{0.9\textwidth}{!}{\input{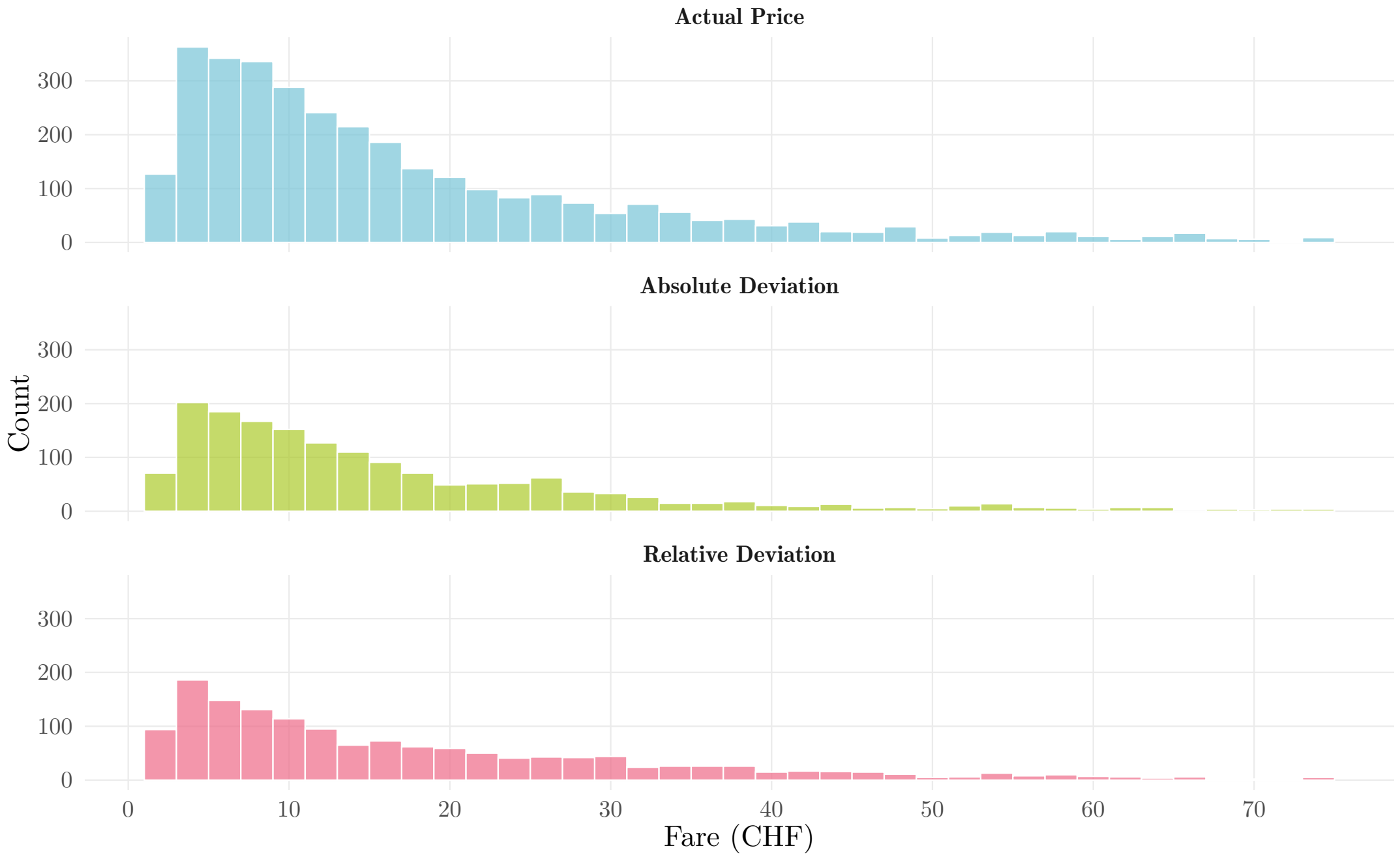}}
		\begin{minipage}{0.9\textwidth}
			\footnotesize \textit{Notes:} Fares above CHF~75 are excluded 
			(approximately 2\% of observations per version: Absolute Amount 2.2\%, 
			Absolute Deviation 1.8\%, Relative Deviation 2.0\%). Binwidth = 
			CHF~2. AbsDev and RelDev fares are converted to absolute CHF amounts 
			using the respondent's actual myRIDE fare as the reference.
		\end{minipage}
	\end{figure}
	
	\begin{figure}[H]
		\centering
		\caption{Predicted Opt-Out Rates by Price Spread and Questionnaire Version}
		\label{fig:optout}
		\resizebox{0.9\textwidth}{!}{
\begin{tikzpicture}[x=1pt,y=1pt]
\definecolor{fillColor}{RGB}{255,255,255}
\path[use as bounding box,fill=fillColor,fill opacity=0.00] (0,0) rectangle (505.89,289.08);
\begin{scope}
\path[clip] (  0.00,  0.00) rectangle (505.89,289.08);
\definecolor{fillColor}{RGB}{255,255,255}

\path[fill=fillColor] (  0.00, -0.00) rectangle (505.89,289.08);
\end{scope}
\begin{scope}
\path[clip] ( 31.71, 67.14) rectangle (500.39,283.58);
\definecolor{drawColor}{gray}{0.92}

\path[draw=drawColor,line width= 0.6pt,line join=round] ( 31.71,104.59) --
	(500.39,104.59);

\path[draw=drawColor,line width= 0.6pt,line join=round] ( 31.71,156.60) --
	(500.39,156.60);

\path[draw=drawColor,line width= 0.6pt,line join=round] ( 31.71,208.61) --
	(500.39,208.61);

\path[draw=drawColor,line width= 0.6pt,line join=round] ( 31.71,260.62) --
	(500.39,260.62);

\path[draw=drawColor,line width= 0.6pt,line join=round] ( 53.02, 67.14) --
	( 53.02,283.58);

\path[draw=drawColor,line width= 0.6pt,line join=round] (159.53, 67.14) --
	(159.53,283.58);

\path[draw=drawColor,line width= 0.6pt,line join=round] (266.05, 67.14) --
	(266.05,283.58);

\path[draw=drawColor,line width= 0.6pt,line join=round] (372.57, 67.14) --
	(372.57,283.58);

\path[draw=drawColor,line width= 0.6pt,line join=round] (479.09, 67.14) --
	(479.09,283.58);
\definecolor{drawColor}{RGB}{119,197,216}

\path[draw=drawColor,line width= 1.1pt,line join=round] ( 53.02,181.38) --
	( 57.28,180.17) --
	( 61.54,178.96) --
	( 65.80,177.76) --
	( 70.06,176.56) --
	( 74.32,175.36) --
	( 78.58,174.17) --
	( 82.84,172.97) --
	( 87.10,171.79) --
	( 91.36,170.60) --
	( 95.62,169.42) --
	( 99.88,168.24) --
	(104.14,167.07) --
	(108.40,165.90) --
	(112.67,164.73) --
	(116.93,163.56) --
	(121.19,162.40) --
	(125.45,161.24) --
	(129.71,160.09) --
	(133.97,158.94) --
	(138.23,157.79) --
	(142.49,156.65) --
	(146.75,155.50) --
	(151.01,154.37) --
	(155.27,153.23) --
	(159.53,152.10) --
	(163.79,150.97) --
	(168.05,149.85) --
	(172.32,148.73) --
	(176.58,147.61) --
	(180.84,146.49) --
	(185.10,145.38) --
	(189.36,144.27) --
	(193.62,143.17) --
	(197.88,142.07) --
	(202.14,140.97) --
	(206.40,139.88) --
	(210.66,138.79) --
	(214.92,137.70) --
	(219.18,136.62) --
	(223.44,135.54) --
	(227.70,134.46) --
	(231.97,133.39) --
	(236.23,132.32) --
	(240.49,131.25) --
	(244.75,130.19) --
	(249.01,129.13) --
	(253.27,128.07) --
	(257.53,127.02) --
	(261.79,125.97) --
	(266.05,124.92) --
	(270.31,123.88) --
	(274.57,122.84) --
	(278.83,121.80) --
	(283.09,120.77) --
	(287.35,119.74) --
	(291.62,118.71) --
	(295.88,117.69) --
	(300.14,116.67) --
	(304.40,115.66) --
	(308.66,114.65) --
	(312.92,113.64) --
	(317.18,112.63) --
	(321.44,111.63) --
	(325.70,110.63) --
	(329.96,109.64) --
	(334.22,108.65) --
	(338.48,107.66) --
	(342.74,106.68) --
	(347.00,105.69) --
	(351.27,104.72) --
	(355.53,103.74) --
	(359.79,102.77) --
	(364.05,101.81) --
	(368.31,100.84) --
	(372.57, 99.88) --
	(376.83, 98.92) --
	(381.09, 97.97) --
	(385.35, 97.02) --
	(389.61, 96.07) --
	(393.87, 95.13) --
	(398.13, 94.19) --
	(402.39, 93.25) --
	(406.65, 92.32) --
	(410.92, 91.39) --
	(415.18, 90.47) --
	(419.44, 89.54) --
	(423.70, 88.62) --
	(427.96, 87.71) --
	(432.22, 86.79) --
	(436.48, 85.89) --
	(440.74, 84.98) --
	(445.00, 84.08) --
	(449.26, 83.18) --
	(453.52, 82.28) --
	(457.78, 81.39) --
	(462.04, 80.50) --
	(466.30, 79.61) --
	(470.57, 78.73) --
	(474.83, 77.85) --
	(479.09, 76.98);
\definecolor{drawColor}{RGB}{238,106,135}

\path[draw=drawColor,line width= 1.1pt,line join=round] ( 53.02,273.74) --
	( 57.28,271.96) --
	( 61.54,270.18) --
	( 65.80,268.40) --
	( 70.06,266.63) --
	( 74.32,264.86) --
	( 78.58,263.10) --
	( 82.84,261.34) --
	( 87.10,259.58) --
	( 91.36,257.83) --
	( 95.62,256.08) --
	( 99.88,254.34) --
	(104.14,252.60) --
	(108.40,250.86) --
	(112.67,249.13) --
	(116.93,247.40) --
	(121.19,245.67) --
	(125.45,243.95) --
	(129.71,242.24) --
	(133.97,240.53) --
	(138.23,238.82) --
	(142.49,237.12) --
	(146.75,235.42) --
	(151.01,233.72) --
	(155.27,232.03) --
	(159.53,230.35) --
	(163.79,228.67) --
	(168.05,226.99) --
	(172.32,225.32) --
	(176.58,223.65) --
	(180.84,221.99) --
	(185.10,220.33) --
	(189.36,218.68) --
	(193.62,217.03) --
	(197.88,215.39) --
	(202.14,213.75) --
	(206.40,212.11) --
	(210.66,210.48) --
	(214.92,208.86) --
	(219.18,207.24) --
	(223.44,205.62) --
	(227.70,204.01) --
	(231.97,202.41) --
	(236.23,200.81) --
	(240.49,199.21) --
	(244.75,197.62) --
	(249.01,196.03) --
	(253.27,194.45) --
	(257.53,192.88) --
	(261.79,191.31) --
	(266.05,189.74) --
	(270.31,188.18) --
	(274.57,186.63) --
	(278.83,185.07) --
	(283.09,183.53) --
	(287.35,181.99) --
	(291.62,180.45) --
	(295.88,178.92) --
	(300.14,177.40) --
	(304.40,175.88) --
	(308.66,174.37) --
	(312.92,172.86) --
	(317.18,171.35) --
	(321.44,169.85) --
	(325.70,168.36) --
	(329.96,166.87) --
	(334.22,165.39) --
	(338.48,163.91) --
	(342.74,162.44) --
	(347.00,160.97) --
	(351.27,159.51) --
	(355.53,158.06) --
	(359.79,156.61) --
	(364.05,155.16) --
	(368.31,153.72) --
	(372.57,152.29) --
	(376.83,150.86) --
	(381.09,149.43) --
	(385.35,148.01) --
	(389.61,146.60) --
	(393.87,145.19) --
	(398.13,143.79) --
	(402.39,142.39) --
	(406.65,141.00) --
	(410.92,139.62) --
	(415.18,138.24) --
	(419.44,136.86) --
	(423.70,135.49) --
	(427.96,134.13) --
	(432.22,132.77) --
	(436.48,131.42) --
	(440.74,130.07) --
	(445.00,128.73) --
	(449.26,127.39) --
	(453.52,126.06) --
	(457.78,124.74) --
	(462.04,123.42) --
	(466.30,122.10) --
	(470.57,120.79) --
	(474.83,119.49) --
	(479.09,118.19);
\end{scope}
\begin{scope}
\path[clip] (  0.00,  0.00) rectangle (505.89,289.08);
\definecolor{drawColor}{gray}{0.30}

\node[text=drawColor,anchor=base east,inner sep=0pt, outer sep=0pt, scale=  0.88] at ( 26.76,101.56) {20};

\node[text=drawColor,anchor=base east,inner sep=0pt, outer sep=0pt, scale=  0.88] at ( 26.76,153.57) {25};

\node[text=drawColor,anchor=base east,inner sep=0pt, outer sep=0pt, scale=  0.88] at ( 26.76,205.58) {30};

\node[text=drawColor,anchor=base east,inner sep=0pt, outer sep=0pt, scale=  0.88] at ( 26.76,257.59) {35};
\end{scope}
\begin{scope}
\path[clip] (  0.00,  0.00) rectangle (505.89,289.08);
\definecolor{drawColor}{gray}{0.30}

\node[text=drawColor,anchor=base,inner sep=0pt, outer sep=0pt, scale=  0.88] at ( 53.02, 56.13) {0.00};

\node[text=drawColor,anchor=base,inner sep=0pt, outer sep=0pt, scale=  0.88] at (159.53, 56.13) {0.25};

\node[text=drawColor,anchor=base,inner sep=0pt, outer sep=0pt, scale=  0.88] at (266.05, 56.13) {0.50};

\node[text=drawColor,anchor=base,inner sep=0pt, outer sep=0pt, scale=  0.88] at (372.57, 56.13) {0.75};

\node[text=drawColor,anchor=base,inner sep=0pt, outer sep=0pt, scale=  0.88] at (479.09, 56.13) {1.00};
\end{scope}
\begin{scope}
\path[clip] (  0.00,  0.00) rectangle (505.89,289.08);
\definecolor{drawColor}{RGB}{0,0,0}

\node[text=drawColor,anchor=base,inner sep=0pt, outer sep=0pt, scale=  1.10] at (266.05, 44.09) {Price differential between options (share of mean price)};
\end{scope}
\begin{scope}
\path[clip] (  0.00,  0.00) rectangle (505.89,289.08);
\definecolor{drawColor}{RGB}{0,0,0}

\node[text=drawColor,rotate= 90.00,anchor=base,inner sep=0pt, outer sep=0pt, scale=  1.10] at ( 13.08,175.36) {Predicted opt-out rate (\%)};
\end{scope}
\begin{scope}
\path[clip] (  0.00,  0.00) rectangle (505.89,289.08);
\definecolor{drawColor}{RGB}{0,0,0}

\node[text=drawColor,anchor=base west,inner sep=0pt, outer sep=0pt, scale=  1.10] at (163.01, 14.44) {Version};
\end{scope}
\begin{scope}
\path[clip] (  0.00,  0.00) rectangle (505.89,289.08);
\definecolor{drawColor}{RGB}{119,197,216}

\path[draw=drawColor,line width= 1.1pt,line join=round] (205.48, 18.23) -- (217.04, 18.23);
\end{scope}
\begin{scope}
\path[clip] (  0.00,  0.00) rectangle (505.89,289.08);
\definecolor{drawColor}{RGB}{238,106,135}

\path[draw=drawColor,line width= 1.1pt,line join=round] (278.98, 18.23) -- (290.54, 18.23);
\end{scope}
\begin{scope}
\path[clip] (  0.00,  0.00) rectangle (505.89,289.08);
\definecolor{drawColor}{RGB}{0,0,0}

\node[text=drawColor,anchor=base west,inner sep=0pt, outer sep=0pt, scale=  0.88] at (223.99, 15.20) {Actual Price};
\end{scope}
\begin{scope}
\path[clip] (  0.00,  0.00) rectangle (505.89,289.08);
\definecolor{drawColor}{RGB}{0,0,0}

\node[text=drawColor,anchor=base west,inner sep=0pt, outer sep=0pt, scale=  0.88] at (297.49, 15.20) {Relative Deviation};
\end{scope}
\end{tikzpicture}}
		\begin{minipage}{0.9\textwidth}
			\footnotesize \textit{Notes:} Predicted opt-out rates from a 
			logistic regression of the opt-out indicator on the relative 
			price spread $\Delta p$, questionnaire version (Absolute Amount 
			and Relative Deviation), and their interaction. A likelihood 
			ratio test of parallel gradients fails to reject the null 
			hypothesis ($\chi^2 = 0.568$, df~$= 1$, $p = 0.451$), 
			indicating no differential selection into the connection 
			choice sample across versions.
		\end{minipage}
	\end{figure}
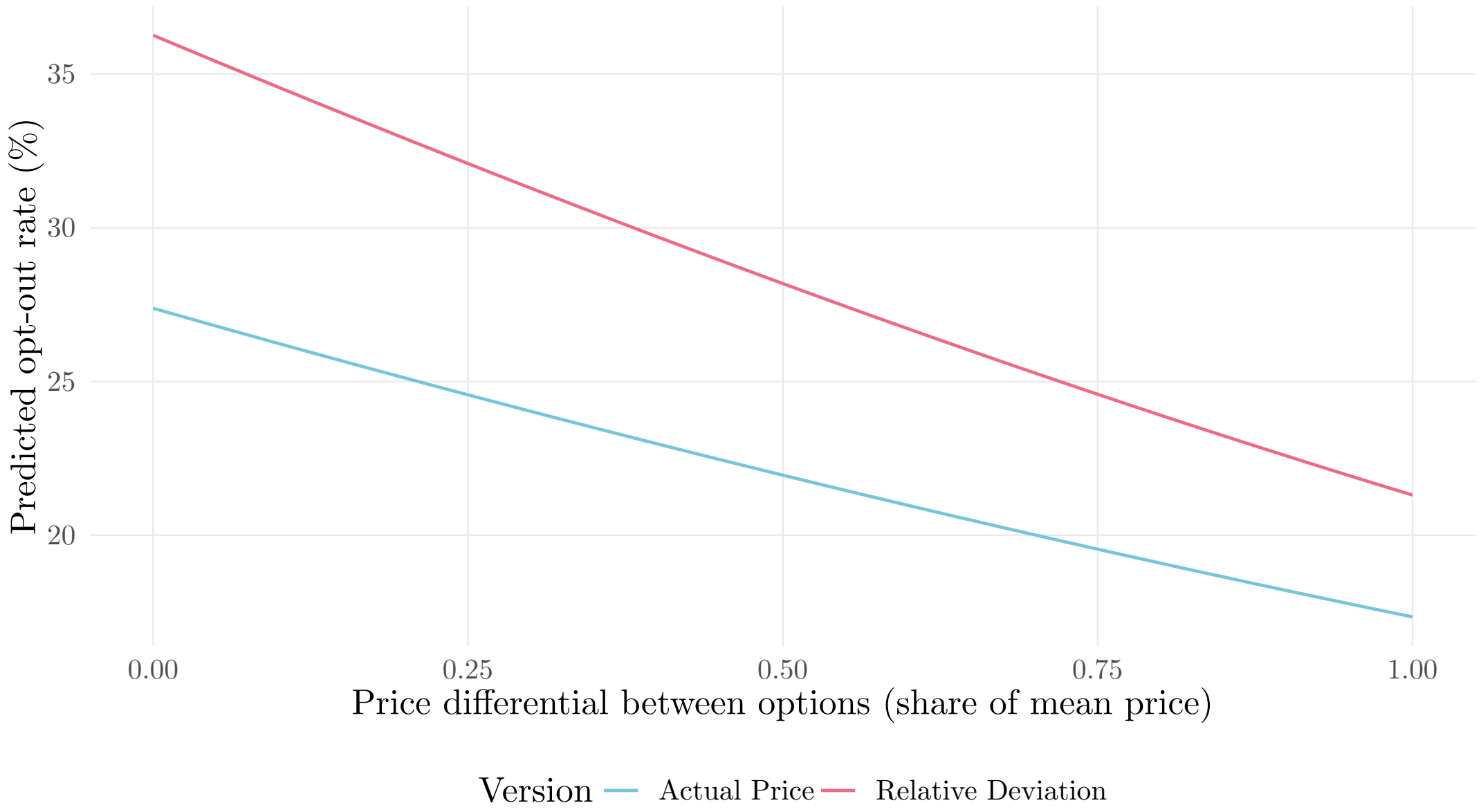

\end{appendix}
\end{document}